\documentclass[aps,prd,groupedaddress,showpacs,nofootinbib,twocolumn]{revtex4}
\usepackage{graphicx,amssymb}

\begin{document}

\title{Can MONDian vector theories explain the cosmic speed up ?}

\author{Vincenzo F. Cardone$^{1,2}$\footnote{{\tt winnyenodrac@gmail.com}}, 
Ninfa Radicella$^{2,3}$\footnote{{\tt ninfa.radicella@polito.it}}}

\affiliation{$^1$Dipartimento di Fisica Generale "Amedeo Avogadro", Via Pietro
Giuria 1, 10125 - Torino, Italy}

\affiliation{$^2$ I.N.F.N. - Sezione di Torino, Via Pietro
Giuria 1, 10125 - Torino, Italy}

\affiliation{$^3$Dipartimento di Fisica, Politecnico di Torino, 
Corso Duca degli Abruzzi 24, 10129 - Torino, Italy}

\begin{abstract}

Generalized Einstein\,-\,Aether vector field models have been shown
to provide, in the weak field regime, modifications to gravity which 
can be reconciled with the successfull MOND proposal. Very little is 
known, however, on the function ${\cal{F}}({\cal{K}})$ defining the 
vector field Lagrangian so that an analysis of the viability of such 
theories at the cosmological scales has never been performed. As a first
step along this route, we rely on the relation between ${\cal{F}}({\cal{K}})$ 
and the MOND interpolating function $\mu(a/a_0)$ to assign the vector
field Lagrangian thus obtaining what we refer to as {\it MONDian vector models}. 
Since they are able by construction to recover the MOND successes on galaxy
scales, we investigate whether they can also drive the observed accelerated
expansion by fitting the models to the Type Ia Supernovae data. Should be this 
the case, we have a unified framework where both dark energy and dark matter can
be seen as different manifestations of a single vector field. It turns out that
both MONDian vector models are able to well fit the low redshift data on Type Ia
Supernovae, while some tension could be present in the high $z$ regime.

\end{abstract}


\maketitle

\section{Introduction}

The old standard cosmological model of a universe full of 
baryon matter only and regulated by the laws of Einstein
General Relativity provided a remarkable picture to consistently
explain the background evolution of the universe from the initial
singularity up to the present day. Unfortunately, two observational
facts then putted in serious trouble this elegant scenario. 
On the one hand, it turned out that the gravitational field measured 
on different scales may not be reconciled with what is inferred from
baryonic matter only, the most famous example being represented by the 
flat rotation curves of spiral galaxies \cite{SR01}. On the other hand, 
recent data on the Hubble diagram of Type Ia Supernovae (hereafter SNeIa)
provided the first evidence for an accelerated expansion \cite{SNeIaFirst}. 
Surprising as it was, the cosmic speed up has been then confirmed by 
updated SNeIa data \cite{SNeIaSec,SNeIaHighZ,SNLS,ESSENCE,D07}, the anisotropy 
spectrum of the Cosmic Microwave Background Radiation (CMBR) as measured
by both balloon \cite{CMBR} and satellites \cite{wmap,WMAP5} experiments, and the
data on the large scale clustering of galaxies \cite{LSS} as estimated from 
large spectroscopic galaxy surveys.

Both these problems have been traditionally solved by adding new ingredients 
to the cosmic pie, while leaving unchanged the underlying theory. Cold dark matter
(CDM) has been then invoked to fill the gap between the baryon mass (constrained
by the Big Bang nucleosynthesis) and what is needed to reproduce the observed
gravitational field. Moreover, this mysterious component must not interact 
with anything but gravitationally so to have escaped any detection notwithstanding
the incredible efforts spent up to now to find direct evidence for any dark particle. At 
the cosmological scales, dark matter still behaves as normal matter so that a further
component is needed in order to drive the accelerated expansion. This new actor on the
scene must, moreover, have an unusual negative pressure and not cluster on galactic 
and cluster scales in order to not oppose gravitational clustering. Referred to as 
dark energy, the nature and nurture of this new term is still debated with the 
cosmological constant $\Lambda$ \cite{Lambda} being the simplest candidate. Although 
added by hand {\it ad hoc}, the cosmological model comprising both of them and known
as $\Lambda$CDM is able to fit extremely well the full dataset at disposal \cite{LCDMtest}
thus worthing the name of {\it concordace model}. Notwithstanding this remarkable success,
the $\Lambda$CDM is theoretically unappealing because of several well known shortcomings 
which motivated the search for other dark energy candidates such as quintessence 
scalar fields \cite{QuintRev} and modified gravity theories, either introducing higher 
dimensions \cite{DGP} or correting the gravitational Lagrangian \cite{stbook,FOG}.

It is worth noting that both dark matter and dark energy can be avoided if one
accepts that the problem is not with what is missing, but rather with how we 
describe gravity. From this point of view, all the evidences for dark matter and
dark energy should be rather seen as evidence that something is wrong with the 
Newton\,-\,Einstein theory of gravity. As a next step, one has therefore to
look for a way to modify gravity at both the galactic and cosmological scales
possibly finding a unified mechanism explaining phenomena taking place on
very different scales such as the flatness of rotation curves and the 
acceleration of the universe expansion. 

Generalized Einstein\,-\,Aether theories are one of these proposed mechanisms 
based on the introduction of dynamical timelike vector field with noncanonical
kinetic terms. Such a proposal is not fully new, but builds upon the extensive
analysis of standard Einstein\,-\,Aether theories \cite{AE} as phenomenological
probes of Lorenz violation in quantum gravity. Indeed, the nonvanishing expectation
value of the Aether field will dynamically select a preferred frame in the spacetime, 
namely the one in which the time coordinate basis vector $\partial_t$ is aligned with 
the direction of the vector field, thus leading to violation of the Lorenz and gauge
invariance. The interest in these models was then renewed after the demonstration 
\cite{ZFS06} that TeVeS, i.e. the fully relativistic and complete MOND theory 
\cite{TeVeS}, can be reformulated as a generalized Einstein\,-\,Aether theory. Later, it 
has been shown that it is possible to get a MOND\,-\,like behaviour in the weak field
regime of vector models \cite{ZFS}. Motivated by these considerations, we 
explore here the cosmological viability of these theories by suitably setting the 
vector field Lagrangian in such a way that the successfull MOND phenomenology is 
recovered in the Newtonian limit.

The structure of the paper is as follows. The basic equations and
properties of the vector models are described in Sect.\,II where we 
also explain how we choose the vector field Lagragian. Sect.\,III is 
devoted to a detailed description of the statistical methodology and 
the data used to test the models in the low redhift regime with a 
discussion of the results obtained from the likelihood analysis. A
similar test is then presented and discussed in Sect.\,IV where we add
high redhift data to probe this complementary redshift range. Conclusions 
and perspectives are finally given in Sect.\,V.

\section{The model}

A general action for a vector field ${\bf A}$ coupled to gravity
can be written as

\begin{equation}
S = \int{d^4x \sqrt{-g} \left [ \frac{R}{16 \pi G_N}
+ {\cal{L}}(g, A) \right ]} + S_M
\label{eq: action}
\end{equation}
where $G_N$ is the Newton gravitational constant, ${\bf g}$ 
the metric (with the signature -, +, +, +), $R$ the Ricci scalar 
and $S_M$ the matter action. The vector field Lagrangian ${\cal{L}}$ may be
a whatever covariant and local function, but, 
following \cite{ZFS}, we will only consider the case\,:

\begin{equation}
{\cal{L}}(g, A) = \frac{M^2}{16 \pi G_N} {\cal{F}}({\cal{K}})
+ \frac{\lambda (A^{\alpha} A^{\alpha} + 1)}{16 \pi G_N} 
\label{eq: lagr}
\end{equation}
where ${\cal{F}}$ is a generic function of 

\begin{equation}
{\cal{K}} = M^{-2} {\cal{K}}^{\alpha \beta}_{\ \ \gamma \sigma}
\nabla_{\alpha} A^{\gamma} \nabla_{\beta} A^{\sigma}
\label{eq: defkappa}
\end{equation}

\begin{equation}
{\cal{K}}^{\alpha \beta}_{\ \ \gamma \sigma} = 
c_1 g^{\alpha \beta} g_{\gamma \sigma} +
c_2 \delta^{\alpha}_{\ \gamma} \delta^{\beta}_{\ \sigma} + 
c_3 \delta^{\alpha}_{\ \sigma} \delta^{\beta}_{\ \gamma} \ .
\label{eq: defkappatens}
\end{equation}
It is worth noticing that ${\cal{K}}$ represents  
all the possible canonical kinetic terms that we introduce
in the vector Lagrangian through the generic function 
${\cal{F}}({\cal{K}})$ thus allowing to have also a noncanonical
contribution by these terms. The $c_i$ quantities are 
dimensionless constants, while $M$ is a scaling mass 
parameter and $\lambda$ a nondynamical Lagrange multiplier 
with dimensions of mass squared. 

It is possible to show \cite{CL04,HBH05,ZFS} that the
Einstein field equations may still be formally written as

\begin{equation}
G_{\alpha \beta} = \widetilde{T}_{\alpha \beta} 
+ 8 \pi G_N T^{M}_{\alpha \beta}
\label{eq: field}
\end{equation}
with $G_{\alpha \beta}$ and $T^{M}_{\alpha \beta}$
the usual Einstein and matter stress\,-\,energy tensors,
while $\widetilde{T}_{\alpha \beta}$ contains the terms 
related to the vector field and some of its derivatives
(see, e.g., \cite{ZFS} for its full expression), while the
equation of motion for the vector field reads\,:

\begin{equation}
\nabla_{\alpha}({\cal{F}}^{\prime} J^{\alpha}_{\beta})
+ {\cal{F}}^{\prime} y_{\beta} = 2 \lambda A^{\gamma} 
\end{equation}
with ${\cal{F}}^{\prime} = d{\cal{F}}/d{\cal{K}}$ and

\begin{equation}
J^{\alpha}_{\ \sigma} = \left (
{\cal{K}}^{\alpha \beta}_{\ \ \sigma \gamma} +
{\cal{K}}^{\beta \alpha}_{\ \ \gamma \sigma} \right ) 
\nabla_{\beta} A^{\gamma}
\end{equation}
and defined the functional derivative\,:

\begin{equation}
y_{\beta} = \nabla_{\sigma} A^{\eta} \nabla_{\gamma} 
A^{\xi} \frac{\delta \left ( {\cal{K}}^{\sigma \gamma}_{\ \ \eta \xi} \right )}
{\delta A^{\beta}} \ .
\end{equation}
Introducing  the flat Robertson\,-\,Walker metric in Eq.(\ref{eq: field}) 
then gives the modified Friedmann equations which read\,:

\begin{equation}
\left [ 1 - \alpha {\cal{K}}^{1/2} \frac{d}{d{\cal{K}}}
\left ( \frac{{\cal{F}}}{{\cal{K}}^{1/2}} \right ) \right ] 
H^2 = \frac{8 \pi G_N}{3} \rho \ ,
\label{eq: f1}
\end{equation}

\begin{equation}
\frac{d}{dt} \left [ \left ( \alpha \frac{d{\cal{F}}}{d{\cal{K}}} 
- 2 \right ) H \right ] = 8 \pi G_N (\rho + p) \ ,
\label{eq: f2}
\end{equation}
where, because of the metric symmetry, it is\,:

\begin{equation}
{\cal{K}} = 3 \alpha \frac{H^2}{M^2} = \frac{3 \alpha}{\varepsilon^2}
\frac{H^2}{H_0^2} \ .
\label{eq: kappacosmo}
\end{equation}
Here, we have defined two new parameters, namely $\varepsilon = M/H_0$
and $\alpha = c_1 + 3 c_2 + c_3$, while $(\rho, p)$ 
are the total  energy density and pressure of the source terms (matter, radiation,
neutrinos, etc.) and $H = \dot{a}/a$ is the usual Hubble parameter. 
Hereafter, we will denote derivatives with respect to $t$ with a dot
and to ${\cal{K}}$ with a prime, and use a subscript $''0''$ to 
label present day quantities.

It is convenient to rearrange the above equations in a different
form. To this end, we first solve Eq.(\ref{eq: f1}) to get\,:

\begin{equation}
\alpha {\cal{F}}^{\prime}({\cal{K}}) = 
1 - \frac{\rho(z)}{\rho_{crit} E^2(z)} + 
\frac{\alpha {\cal{F}}({\cal{K}})}{2 {\cal{K}}} 
\label{eq: f1new}
\end{equation}
with $E(z) = H(z)/H_0$ the dimensionless Hubble parameter. 
Inserting Eq.(\ref{eq: f1new}) into Eq.(\ref{eq: f2})
and using

\begin{displaymath}
\frac{d}{dt} = -(1 + z) H_0 E(z) \frac{d}{dz}
\end{displaymath}
to change variable, one finally gets\,:

\begin{equation}
\frac{d}{dz} \left \{ \left [
\frac{\varepsilon^2 {\cal{F}}({\cal{K}})}{6 E^2(z)}
- \frac{\rho/\rho_{crit}}{E^2(z)} - 1 \right ]
E^2(z) \right \} = - \frac{3 (\rho + p)/\rho_{crit}}{(1 + z) E(z)}
\label{eq: masteq}
\end{equation}
with $\rho_{crit} = 3H_0^2/8 \pi G_N$ the present day 
critical density. Let us now assume that the universe is filled 
by a matter\,-\,like term and radiation. Since the vector field 
is only coupled to gravity, the continuity equation

\begin{displaymath}
\dot{\rho} + 3 H (\rho + p) = 0
\end{displaymath}
still holds for both matter and radiation and we can
therefore write the rhs of Eq.(\ref{eq: masteq}) as

\begin{equation}
\frac{3(\rho + p)/\rho_{crit}}{(1 + z) E(z)} = 
\frac{\left [ 3 \Omega_M + 4 \Omega_r (1 + z) \right ]
(1 + z)^2}{E(z)}
\label{eq: rhsmaster}
\end{equation}
with $\Omega_i = \rho_i(z = 0)/\rho_{crit}$ the present 
day density parameter of the i\,-\,th component.

Eqs.(\ref{eq: f1new})\,-\,(\ref{eq: masteq}) clearly show that
a key role in determining the cosmic evolution is played by 
the functional expression adopted for ${\cal{F}}({\cal{K}})$. 
A guide to the choice of this quantity is provided by the observation
that, in the nonrelativistic regime, the field equations reduce to\,:

\begin{equation}
\nabla \cdot \left [ \left ( 2 + c_1 \frac{d{\cal{F}}}{d{\cal{K}}}
\right ) \nabla \Phi \right ] = 8 \pi G_N \rho  
\label{eq: modpoisson}
\end{equation}
with 

\begin{equation}
{\cal{K}} = -c_1 \frac{|\nabla \Phi|^2}{M^2} \ , 
\label{eq: kappaphi}
\end{equation}
being $\Phi$ the gravitational potential sourced by 
the density distribution $\rho$. By using the above 
expression, we get\,:

\begin{equation}
\nabla \cdot \left [ \mu(|\nabla \Phi|/M) \nabla \Phi \right ]
= 8 \pi G_N \rho
\label{eq: mondpoisson}
\end{equation}
which is the modified Poisson equation for the 
MOND theory\footnote{Actually, such a modified Poisson
equation was proposed for MOND in the framework of the so
called {\it AQUAL} theory \cite{BM84}, one of the first
attempts to work out a relativistic MOND theory. AQUAL was later
abandoned since it turned out to be unable to account for lensing
data without cold dark matter thus being in contrast with the original
MOND philosophy.}, provided we identify the MOND interpolating function $\mu(a/a_0)$ with

\begin{equation}
\mu(a/a_0) = \mu(\sqrt{{\cal{K}}}) = 2 + 
c_1 \frac{d{\cal{F}}}{d{\cal{K}}} \ .
\label{eq: muvsfk}
\end{equation}
Note that, by this position, the MOND acceleration scale
$a_0$ turns out to be related to the mass scale $M$ as

\begin{equation}
a_0 = \frac{\varepsilon c H_0}{\sqrt{-c_1}} \ \ 
\ \ \iff \ \ c_1 = - \left ( \frac{\varepsilon c H_0}{a_0} \right )^2 \ ,
\label{eq: c1a0}
\end{equation}
where we have reintroduced the speed of light $c$. It is 
worth noting that such a model thus provides a natural 
mechanism to explain why one observationally finds that 
$a_0 \sim c H_0$ which now emerges as a consequence
of the local and cosmological phenomena being different 
manifestations of the same underlying theory.

Since we know that MOND makes it possible to fit the flat 
rotation curves of spiral galaxies (see, e.g., \cite{SmG02}
and refs. therein), it is reasonable to assume that a viable
expression for ${\cal{F}}({\cal{K}})$ should lead to the 
same interpolating function $\mu(a/a_0)$ which is successfully
used on local scales. Two such functions are the {\it simple form} \cite{FB05}

\begin{equation}
\mu(x) = \frac{x}{1 + x} \ ,
\label{eq: simplemu}
\end{equation}
and the {\it standard form} \cite{BM84}

\begin{equation}
\mu(x) = \frac{x}{\sqrt{1 + x^2}} \ .
\label{eq: standmu}
\end{equation}
Although Eq.(\ref{eq: standmu}) has been the first proposal to 
be tested with success \cite{SmG02}, recent analyses \cite{FB05,MondFitRC} 
seem to favour Eq.(\ref{eq: simplemu}). 
However, in order to gain further insight on the 
problem of which interpolating function is better motivated, 
we will consider both cases contrasting them against data 
probing radically different scales. 

Inserting alternatively Eqs.(\ref{eq: simplemu}) and (\ref{eq: standmu})
into Eq.(\ref{eq: muvsfk}) gives respectively\,:

\begin{equation}
{\cal{F}}({\cal{K}}) = - \frac{4}{c_1} \left [ 
\sqrt{{\cal{K}}} - \ln{\left ( 1 + \sqrt{{\cal{K}}} \right )}
\right ] \ ,
\label{eq: fksimple}
\end{equation}

\begin{equation}
{\cal{F}}({\cal{K}}) = - \frac{2}{c_1} \left [
{\cal{K}} - \sqrt{{\cal{K}} \left ( 1 + {\cal{K}}
\right )} + \ln{\left ( \sqrt{{\cal{K}}} + 
\sqrt{1 + {\cal{K}}} \right )} \right ] \ .
\label{eq: fkstand}
\end{equation}
Hereafter, we will refer to the models assigned 
by Eqs.(\ref{eq: fksimple}) and (\ref{eq: fkstand}) as the 
{\it simple} and {\it standard} {\it MONDian vector model}
in order to clearly remember that the main ingredient of this kind 
of cosmological theory is the presence of a vector field with
a Lagrangian constructed in such a way to recover the simple or 
standard MOND interpolating function. Both expressions for 
${\cal{F}}({\cal{K}})$ depend on $\sqrt{{\cal{K}}}$ so that it must
be $\alpha > 0$ in the cosmological setting. In the small ${\cal{K}}$
limit, ${\cal{F}}({\cal{K}}) \sim {\cal{K}}$ for both the simple and
standard MONDian vector models so that we can use the perturbative 
analysis in \cite{Lim} for the Einstein\,-\,Aether models to see 
whether the condition $\alpha > 0$ is compatible with the constraints
on the coefficients $c_i$. The requirement that the Hamiltonian for the 
perturbations is positive definite implies $c_1 < 0$, while the 
constraint $(c_1 + c_2 + c_3)/c_1 \ge 0$ has to be set in order to avoid
tachyonic propagation of spin\,-\,0 modes. On the other hand, if we allow
superluminal propagations of both spin\,-\,0 and spin\,-\,2 modes, 
as supported in \cite{spin}, we then get that for $c_2 > 0$ and 

\begin{displaymath}  
-(c_1 + 3c_2) \le c_3 \le -(c_1 + c_2)
\end{displaymath}
it is indeed $\alpha > 0$ so that we can safely consider the 
MONDian vector models without violating any constraint on the 
$c_i$ coefficients.

\subsection{The simple MONDian vector model}

In order to determine the dynamics for this case, we have to
insert Eq.(\ref{eq: fksimple}) into the master equation 
(\ref{eq: masteq}) and solve with respect to $E(z)$. Somewhat
surprisingly, after some algebra, we get\,:

\begin{displaymath}
{\cal{Q}}[z, E(z)] \times \frac{dE}{dz} = 0
\end{displaymath}
with ${\cal{Q}}[z, E(z)]$ an algebraic function of 
redshit $z$, the dimensionless Hubble parameter
$E(z)$ and the constant parameter of the model. Since we
know that the universe is expanding, $dE/dz > 0$ so that we must solve ${\cal{Q}}[z, E(z)] = 0$. 
Rearranging the different terms gives\,:

\begin{equation}
\begin{array}{l}
\eta y^3 + y^2 - \eta \left \{ \kappa [ 1 - 
\ln{(1 + \eta y)}] + [ \Omega_M + \Omega_r (1 + z) ] \right . \\
\left . \times (1 + z)^3 \right \} y + 
\kappa \ln{(1 + \eta y)} =   [ \Omega_M + \Omega_r (1 + z) ]
(1 + z)^3 \} \ ,
\end{array}
\label{eq: simplemast}
\end{equation}
where we have set $y = E(z)$ and defined\,:

\begin{equation}
\eta = \frac{\sqrt{3 \alpha}}{\varepsilon} \ ,
\label{eq: defeta}
\end{equation}

\begin{equation}
\kappa = \frac{2 \varepsilon^2}{3 c_1} = 
- \frac{2 a_0^2}{3 c^2 H_0^2} \simeq -0.01 \ h^{-2} \ .
\label{eq: defkappapar}
\end{equation}
with $h = H_0/(100  / {\rm km/s/Mpc})$. Note that, in 
the rhs of Eq.(\ref{eq: defkappapar}), we have used Eq.(\ref{eq: c1a0}) 
and set $a_0 =  1.2 \times 10^{-10} \ {\rm m/s^2}$ in agreement with
the estimates coming from the MOND fit to the galaxy rotation curves.

In order to reduce the number of parameters of the model, we can insert
Eq.(\ref{eq: fksimple}) into (\ref{eq: f1new}) and evaluate it at $z = 0$.
Remembering that $E(z = 0) = 1$ by definition, we then get\,:

\begin{equation}
\ln{(1 + \eta)} - \frac{\eta}{1 + \eta} = 
\frac{1 - \Omega_M - \Omega_r}{\kappa}
\label{eq: etasolsimple}
\end{equation}
which can be solved numerically for given values of 
$(\Omega_M, \Omega_r, h)$. Actually, since 
$\Omega_r$ is typically set by the CMB temperature, 
the simple MONDian vector model is fully characterized
by only two parameters. This is the same as the concordance
$\Lambda$CDM model where the same parameters $(\Omega_M, h)$ 
have to be assigned in order to compare the theory with the data. 
However, in that scenario, two different ingredients are invoked
in order to explain the dynamics of galaxies and the cosmic
speed up, while here the solution to both problems comes out as
a consequence of the presence of a single vector field.

\subsection{The standard MONDian vector model}

We can repeat the same steps as before to get the master
equation for the case when ${\cal{F}}({\cal{K}})$ is given
by Eq.(\ref{eq: fkstand}). Not surprisingly given the 
similarities of the models, we still get an algebraic 
relation\,:

\begin{equation}
\begin{array}{l}
2 + \kappa \eta^2 + 
\displaystyle{\frac{\kappa \eta}{y \sqrt{1 + \eta^2 y^2}}} 
- \displaystyle{\frac{\kappa \eta^3 y}{\sqrt{1 + \eta^2 y^2}}} = \\
\displaystyle{\frac{2 \left [ \Omega_M + \Omega_r (1 + z) \right ]
(1 + z)^3 + \kappa \ln{\left ( \eta y + \sqrt{1 + \eta^2 y^2} \right )}}
{y^2}}
\end{array}
\label{eq: standmast}
\end{equation}
with $y$, $\eta$ and $\kappa$ defined as above. The relation between 
$\eta$ and the other model parameters may be obtained as before 
and turns out to be\,:

\begin{equation}
\ln{\left ( \eta + \sqrt{1 + \eta^2} \right )}
- \frac{\eta (1 - \eta^2)}{\sqrt{1 + \eta^2}} = 
\frac{2(1 - \Omega_M - \Omega_r)}{\kappa}
\label{eq: etasolstand}
\end{equation}
which has still to be solved numerically. Note that, 
as for the simple case, also the standard MONDian vector
model is characterized by only two quantities, namely
the total matter density parameter $\Omega_M$ and the 
present day scaled Hubble constant $h$.

\section{Testing the low redshift regime}

Any model that aims to describe the evolution of the
universe must be able to reproduce what is indeed observed. This
is particularly true for the models we are considering since we are 
introducing a single vector field to get rid of both dark matter and 
dark energy. Matching the model with observations is also a powerful 
tool to constrain its parameters and, as an interesting byproduct, allows 
us to estimate some quantities common to every model (such as the age of the 
universe $t_0$ and the transition redshift $z_T$) to previous literature values.

As a first test, we are here interested in exploring the 
behaviour of our models in the low redshift regime so that
we start by only considering data probing up to $z \sim 1.5$. To this 
end, we therefore maximize the following likelihood function\,:

\begin{eqnarray}
{\cal{L}}({\bf p}) & \propto  & {\cal{L}}_{SNeIa}({\bf p}) \nonumber \\
~ & \times &  \exp{\left [ - \frac{1}{2}
\left ( \frac{\omega_b^{obs} - \omega_b^{th}}{\delta \omega_b}  \right )^2
\right ]} \nonumber \\
~ & \times & \exp{\left [ - \frac{1}{2} 
\left ( \frac{h_{HST} - h}{\delta h} \right )^2 \right ]}
\label{eq: deflike}
\end{eqnarray}
where {\bf p} denotes the set of model parameters. Before discussing in
detail the term related to the SNeIa data, we concentrate on
the two Gaussian priors. The first one takes into account the 
constraints on the  the physical baryon density $\omega_b = \Omega_b h^2$ with

\begin{displaymath}
\omega_b^{obs} \pm \delta \omega_b = 0.0228 \pm 0.0055
\end{displaymath}
as estimated in \cite{San09} and in agreement with what is
inferred from the abundance of ligth elements \cite{Kirk}. The HST Key
Project \cite{hstkey} has estimated the Hubble constant $H_0$ using a
well calibrated set of local distance scale estimators thus ending up with

\begin{displaymath}
h_{HST} \pm \delta h = 0.720 \pm 0.008
\end{displaymath}
as a final model independent constraint. 

\subsection{The SNeIa data}

Since the first announcement \cite{SNeIaFirst} of the evidence 
of cosmic speed up, the importance of the SNeIa Hubble diagram 
as a probe of the universe background evolution has always been clear. 
It has therefore become a sort of {\it ground zero} for every proposed 
cosmological model to fit the SNeIa data. To this end, one relies on the 
predicted distance modulus\footnote{We use $\ln{x}$ and $\log{x}$
to denote the logarithm base ${\it e}$ and $10$.}\,:

\begin{equation}
\mu_{th}(z, {\bf p}) = 25 + 5 \log{\left [ \frac{c}{H_0} (1 + z) r(z, {\bf p}) \right ]}
\label{eq: defmuth}
\end{equation}
with $r(z)$ the dimensionless comoving distance\,:

\begin{equation}
r(z, {\bf p}) = \int_{0}^{z}{\frac{dz'}{E(z', {\bf p})}} \ .
\label{eq: defrz}
\end{equation}
The likelihood function is then defined as

\begin{eqnarray}
{\cal{L}}_{SNeIa}({\bf p}) & = & 
\frac{1}{(2 \pi)^{N_{SNeIa}/2} |C_{SNeIa}^{-1}|^{1/2}}
\nonumber \\ ~ & \times &
\exp{\left ( - 
\frac{\Delta \mu \cdot C_{SNeIa}^{-1} \cdot \Delta \mu^T}{2} 
\right )} \ ,
\label{eq: deflikesneia}
\end{eqnarray}
where $N_{SNeIa}$ is the total number of SNeIa used, 
$\Delta \mu$ is a $N_{SNeIa}$\,-\,dimensional vector with 
the values of $\mu_{obs}(z_i) - \mu_{th}(z_i)$ and $C_{SNeIa}$
is the $N_{SNeIa} \times N_{SNeIa}$ covariance matrix of the SNeIa data. 
Note that, if we neglect the correlation induced by systematic errors, $C_{SNeIa}$ is a diagonal 
matrix so that Eq.(\ref{eq: deflikesneia}) simplifies to\,:

\begin{equation}
{\cal{L}}_{SNeIa}({\bf p}) \propto \exp{[-\chi_{SNeIa}^2({\bf p})/2]}
\label{eq: likesneiasimple}
\end{equation}
with

\begin{equation}
\chi_{SNeIa}^2({\bf p}) = 
\sum_{i = 1}^{N_{SNeIa}}{\left [ 
\frac{\mu_{obs}(z_i) - \mu_{th}(z_i)}{\sigma_i} \right ]^2}
\label{eq: defchisneia}
\end{equation}
with $\sigma_i$ the error on the observed distance modulus 
$\mu_{obs}(z_i)$ for the i\,-\,th object at redshift $z_i$. As 
input data, we use the Union SNeIa sample assembled in \cite{K08} by
reanalysing with the same pipeline both the recent SNeIa SNLS \cite{SNLS}
and ESSENCE \cite{ESSENCE} samples and older nearby and high redshift 
\cite{SNeIaHighZ} datasets.

\subsection{How many model parameters ?}

As shown by Eqs.(\ref{eq: simplemast}) and (\ref{eq: standmast}),
in order to determine the cosmic dynamics, one has to set the
value of the present day total matter density parameter $\Omega_M$. 
On the other hand, the Gaussian prior on $\omega_b$ in the likelihood
function (\ref{eq: deflike}) and of the distance priors introduced later 
asks for discriminating between the baryons only and total matter 
physical densities. It is therefore worth wondering how $\Omega_M$
and $\Omega_b$ are related. Since our theory reduces to MOND 
in the low energy limit and MOND does not need any cold dark matter
on the galactic scales (hence no other matter than the visible one), 
we could argue that $\Omega_M = \Omega_b$ should hold. On the other hand,
large amounts of missing matter are needed in order MOND to reproduce
the observed phenomenology on cluster scales \cite{MondCl}. It was therefore
postulated \cite{MondNeutrinos} that massive neutrinos (with $m_{\nu} 
\sim 2 \ {\rm eV}$) can play the role of dark mass in galaxy clusters. 
Moreover, solar and atmospheric neutrinos experiments \cite{massnu} 
have shown that the three active neutrinos from the standard model of 
particle physics mix their flavours which is only possible if
they are massive. Finally, it is also worth noting that it has 
been claimed \cite{stnu} that a single sterile neutrino with mass in 
the range 4\,-\,6 eV is better suited to explain the results of the Miniboone
experiment, while a 11 eV sterile neutrino has been indeed advocated
\cite{gary} in order to solve the MOND problems on cluster and
cosmological scales.

Massive neutrinos decoupled at $\sim 1 \ {\rm MeV}$
and since last scattering they have been non relativistic particles 
so that, from a cosmological point of view, they behave exactly as matter. Assuming 
three families of degenerate neutrinos, the total matter density 
parameter will therefore read\,:

\begin{equation}
\Omega_M = \Omega_b + \frac{3 m_{\nu}}{94 h^2 \ {\rm eV}}
\label{eq: omvsobmnu}
\end{equation}
so that the number of model parameters is increased by one 
updating from $(\Omega_M, h)$ to $(\Omega_b, m_{\nu}, h)$.

It is worth stressing, however, that discriminating 
between $\Omega_M$ as a single quantity and $\Omega_M$
as function of $(\Omega_b, m_{\nu}, h)$ is only possible
if the data at hand depend on them separetely. To understand
this point, let us consider the SNeIa Hubble diagram. In 
order to fit this dataset, we just need the dimensionless
Hubble parameter $E(z)$ which is obtained by solving, e.g., 
Eq.(\ref{eq: simplemast}). To this end, we just have to 
set the value of $\Omega_M$ so that all the SNeIa 
likelihood will be a function of $(\Omega_M, h)$ only and 
we cannot set any constraint on $(\Omega_b, m_{\nu})$. That is why
we have added the Gaussian priors on $\omega_b$ and $h$ in order to
make the likelihood explicitly dependent on $\Omega_b$. However, 
since the likelihood is mainly driven by the SNeIa term, one can
forecast that the constraints on $(\Omega_b, m_{\nu})$ will be 
quite weak because of the degeneracy being only partially broken. 

\begin{table}[t]
\begin{center}
\begin{tabular}{ccccc}
\hline
Par $x$ & $\langle x \rangle$ & $x_{med}$ & $68\%$ CL & $95\%$ CL \\
\hline \hline
$\Omega_b$ & 0.13 & 0.12 & (0.05, 0.23) & (0.02, 0.29) \\
$\log{m_{\nu}}$ & 0.24 & 0.40 & (-0.16, 0.55) & (-0.88, 0.63) \\
$h$ & 0.700 & 0.700 & (0.694, 0.706) & (0.689, 0.711) \\
\hline
$\Omega_b$ & 0.15 & 0.13 & (0.07, 0.26) & (0.02, 0.31) \\
$\log{m_{\nu}}$ & -0.09 & 0.35 & (-1.10, 0.53) & (-2.40, 0.64) \\
$h$ & 0.700 & 0.700 & (0.694, 0.706) & (0.688, 0.712) \\
\hline
\end{tabular}
\end{center}
\caption{Summary of the results of the likelihood analysis including
SNeIa and Gaussian priors on $\omega_b$ and $h$. The upper and lower part 
of the table refers to the simple and standard MONDian vector model respectively.}
\end{table}

\subsection{Results}

In order to maximize the likelihood function (\ref{eq: deflike}),
we run a Markov Chain Monte Carlo (MCMC) algorithm to efficiently 
explore the parameters\footnote{Note that, hereafter, we use $\log{m_{\nu}}$ 
rather than $m_{\nu}$ as neutrino mass parameter since this choice allows 
to explore a wider range.} space. We use a single chain with $\sim 100000$
points reduced to $\sim 3100$ after burn in cut and thinning. The 
histograms of the values of each parameter are then used to infer
the median and the mean and the $68$ and $95\%$ confidence ranges
summarized in Table\,I. The best fit parameters (i.e., the set 
maximizing the likelihood) for the simple MONDian vector model turns out to be\,:

\begin{displaymath}
(\Omega_b, \log{m_{\nu}}, h) =  
(0.056, 0.54, 0.70) \ ,
\end{displaymath}
while the standard case gives\,:

\begin{displaymath}
(\Omega_b, \log{m_{\nu}}, h) =  
(0.050, 0.55, 0.70) \ .
\end{displaymath}
For both models we get $\chi^2_{SNeIa}/d.o.f. \simeq 1.02$
thus indicating a very good agreement. Moreover, the physical
baryon density reads $\omega_b = 0.0276$ (0.0246) for the 
simple (standard) model in good agreement with the observed
value well within the $1 \sigma$ error, while both models 
predict values for $h$ very close to the HST Key Project result. 
It is worth noting that a baryon only universe may be safely
excluded since the reduced $\chi^2$ values are of order $2.15$ 
for both cases thus definitevely ruling out models with no massive
netrinos.

A look at Table I shows that the constraints on both $\Omega_b$
and $\log{m_{\nu}}$ are quite weak. Moreover, the best fit values 
of both parameters radically differ from their median values. This
is, however, not an unexpected result. As we have said, should the prior 
on $\omega_b$ be neglected, the model should collapse into a two 
parameter one with $\Omega_M$ replacing the set $(\Omega_b, 
\log{m_{\nu}})$. The prior on $\omega_b$ thus helps
in discriminating between the two, but, from the point of view
of maximizing the likelihood, it is of little help. Indeed, 
the only reliable constraint is on $\Omega_M$ reading\,:

\begin{displaymath}
\langle \Omega_M \rangle = 0.28 \ \ , \ \ 
\Omega_{M,med} = 0.28 \ \ ,
\end{displaymath}

\begin{displaymath}
{\rm 68\% \ CL \ :} \ \ (0.25, 0.31) \ \ ,
\end{displaymath}

\begin{displaymath}
{\rm 95\% \ CL \ :} \ \ (0.23, 0.34) \ \ ,
\end{displaymath}
for the simple MONDian vector model and

\begin{displaymath}
\langle \Omega_M \rangle = 0.28 \ \ , \ \ 
\Omega_{M,med} = 0.28 \ \ ,
\end{displaymath}

\begin{displaymath}
{\rm 68\% \ CL \ :} \ \ (0.25, 0.31) \ \ ,
\end{displaymath}

\begin{displaymath}
{\rm 95\% \ CL \ :} \ \ (0.22, 0.33) \ \ ,
\end{displaymath}
for the standard case. Note that these values are in very 
good agreement with typical estimates from previous analyses 
of comparable datasets \cite{D07,K08,WMAP5}. It is also worth
noting that the results are almost fully independent on the 
functional expression adopted for ${\cal{F}}({\cal{K}})$ which
is an expected consequence of the two models matching each other
in order to fit the same SNeIa Hubble diagram. Provided that
the above constraints on $\Omega_M$ are met, we can choose whatever
value of $\Omega_b$ (for a given $h$) and then find a corresponding 
$\log{m_{\nu}}$ value giving rise to a model with a given likelihood
value, i.e. ${\cal{L}}$ depends only on $(\Omega_M, h)$. As a 
consequence, it is therefore not surprising that the constraints
on $(\Omega_b, \log{m_{\nu}})$ are so weak.

With this caveat in mind, it is nevertheless interesting to look
at the neutrino mass. Converting the constraints on $\log{m_{\nu}}$
into constraints on $m_{\nu}$ (in eV), we get\,:

\begin{displaymath}
\langle m_{\nu} \rangle = 2.3 \ \ , \ \ 
m_{\nu,med} = 2.5 \ \ ,
\end{displaymath}

\begin{displaymath}
{\rm 68\% \ CL \ :} \ \ (0.7, 3.6) \ \ ,
\end{displaymath}

\begin{displaymath}
{\rm 95\% \ CL \ :} \ \ (0.1, 4.2) \ \ ,
\end{displaymath}
for the simple MONDian vector model and

\begin{displaymath}
\langle m_{\nu} \rangle = 2.0 \ \ , \ \ 
m_{\nu,med} = 2.2 \ \ ,
\end{displaymath}

\begin{displaymath}
{\rm 68\% \ CL \ :} \ \ (0.1, 3.4) \ \ ,
\end{displaymath}

\begin{displaymath}
{\rm 95\% \ CL \ :} \ \ (0.0, 4.4) \ \ ,
\end{displaymath}
for the standard case. As yet stated, atmospheric and
solar neutrino experiments have shown that the three
families of standard model neutrinos are massive, but 
they are unable to put any constraints on their exact 
masses being only sensitive to mass squared differences. 
An upper limit on the mass may instead be set from the
study of the tritium $\beta$ decay. By this method, the 
Mainz\,-\,Troitz experiment \cite{MT05} was able to 
find $m_{\nu} \le 2.2 \ {\rm eV}$. The median $m_{\nu}$
values quoted above are smaller than this upper limit, 
while the $68$ and $95\%$ confidence ranges do indeed
suggest that it is possible to fit the data equally well
with still lighter neutrinos. On the other hand, it is
worth stressing that such estimates rely on our 
assumption that three degenerate massive neutrinos 
are present so that their total density parameter 
reads $\Omega_{\nu} = 3 m_{\nu}/94 h^2$. It has, 
however, been claimed \cite{stnu} that a single sterile 
neutrino with mass in the range 4\,-\,6 eV is better suited
to explain the results of the Miniboone experiment. Should 
this be the case, the constraints above should be multiplied
by three thus giving a single sterile neutrino with mass in the 
$95\%$ confidence range $(0.3, 12.6) \ {\rm eV}$. It is worth
noting that a 11 eV sterile neutrino has been indeed advocated
\cite{gary} in order to solve the MOND problems on cluster and
cosmological scales. It is therefore tempting to investigate
whether the results in \cite{gary} still hold for our vector 
models since on cluster scales both theories reduce to MOND.

The MCMC algorithm makes also possible to infer constraints on
some interesting derived quantities. To this end, one just has
to evaluate a given function $f({\bf p})$ for each point of the 
chain and then use the values thus obtained as yet done for
the parameters ${\bf p}$. Table\,II summarizes the results for 
the present day deceleration parameter $q_0$, the transition
redshift $z_T$ obatined by solving $q(z_T) = 0$ and the 
age of the universe $t_0$ estimated as

\begin{equation}
t_0 = t_H \int_{0}^{\infty}{\frac{dz}{(1 + z) E(z)}} 
\label{eq: deftz}
\end{equation}
with $t_H = 9.78 h^{-1} \ {\rm Gyr}$ the Hubble time. 

\begin{table}[t]
\begin{center}
\begin{tabular}{ccccc}
\hline
Par $x$ & $\langle x \rangle$ & $x_{med}$ & $68\%$ CL & $95\%$ CL \\
\hline \hline
$q_0$ & -0.57 & -0.58 & (-0.62, -0.53) & (-0.65, -0.49) \\
$z_T$  & 0.73 & 0.72 & (0.65, 0.80) & (0.58, 0.88) \\
$t_0$ & 13.71 & 13.70 & (13.43, 14.00) & (13.18, 14.31) \\
\hline
$q_0$ & -0.58 & -0.58 & (-0.62, -0.53) & (-0.65, -0.49) \\
$z_T$  & 0.73 & 0.72 & (0.64, 0.81) & (0.57, 0.88) \\
$t_0$ & 13.72 & 13.71 & (13.42, 14.02) & (13.16, 14.29) \\
\hline
\end{tabular}
\end{center}
\caption{Constraints on derived quantities ($t_0$ in Gyr) from 
the chains obtained fitting the SNeIa with Gaussian priors on $\omega_b$ and $h$. 
The upper and lower part of the table refers to the simple and standard 
MONDian vector model respectively.}
\end{table}

As a first issue, let us consider the value of $q_0$. Its 
estimate is typically model dependent since it comes out as 
a derived quantity given a model parametrization. In order to
escape this problem, one may resort to cosmographic analyses based
only on Taylor expanding the scale factor. Using this approach, 
Catto\"en and Visser \cite{CV08} have found values between 
$q_0 = -0.48 \pm 0.17$ and  $q_0 = -0.75 \pm 0.17$ depending on the 
details of the  method used to fit the SNLS dataset. A similar analysis 
but using the GRBs as distance indicators allowed Capozziello 
and Izzo \cite{CI08} to find values between $q_0 = -0.94 \pm 0.30$
and $q_0 = -0.39 \pm 0.11$ still in accordance with our
estimates. A different approach has been instead adopted 
by Elgar$\o$y and Multam\"aki \cite{EM06} advocating a model independent 
parametrization of $q(z)$. Depending on the SNeIa sample used and 
the parametrization adopted, their best fit values for $q_0$ 
range between -0.29 and -1.1 in good agreement with the
estimates in Table II. 

On the contrary, there is some conflict with the 
previous estimates of the transition redshift $z_T$. For instance, 
using the Gold SNeIa sample  and linearly expanding $q(z)$, 
Riess et al. \cite{SNeIaHighZ} found $z_T = 0.46 \pm 0.13$ 
in agreement with the updated result $z_T = 0.49_{-0.07}^{+0.14}$ 
obtained by Cuhna \cite{Cu09} using the Union sample. Although there 
is a possible marginal agreement within the $95\%$ 
confidence range, we nevertheless consider this 
unsatisfactory result not a serious flaw of our models 
since the estimate of $z_T$ is strongly model dependent 
so that it is not possible to decide whether the 
disageement is with the data or with the fiducial model
used to fit the data. 

Finally, we note that the age of the universe is in agreement 
with previous estimates in the literature. Fitting the WMAP5 data 
and the SNLS SNeIa sample with a prior on the acoustic
peak parameter gives \cite{WMAP5} $t_0 = 13.73 \pm 0.12 \ {\rm Gyr}$ in
almost perfect agreement with the results in Table II. Moreover, the
shape of the colour\,-\,magnitude diagram of globular clusters provides 
a model independent estimate, namely $t_0 =  12.6_{-2.6}^{+3.4} \ {\rm Gyr}$ 
\cite{Krauss}, still in considerable good agreement with our values.

Summarizing, the very good fits to the SNeIa data and the agreement 
between observed and predicted derived quantities make us confident 
that both the simple and standard MONDian vector models successfully 
reproduces the data thus being viable alternatives to the usual dark energy 
models in the low redshift regime.

\section{Probing the high redshift regime}

The SNeIa Hubble diagram and the Gaussian priors on $\omega_b$ 
and $h$ allow us to test the behaviour of the MONDian vector 
models only over the redshift range probed by these data. Considering
that the farthest SN has redshift $z \simeq 1.6$, it is worth 
wondering whether the models work well for higher $z$. To answer 
this question, we change the likelihood function adding a further
term\,:

\begin{equation}
{\cal{L}}({\bf p}) \propto {\cal{L}}_{low}({\bf p}) \ 
\times \ {\cal{L}}_{dp}({\bf p})
\label{eq: deflikenew}
\end{equation}
with ${\cal{L}}_{low}({\bf p})$ the likelihood term related
to low redshift data given by Eq.(\ref{eq: deflike}), while
${\cal{L}}_{dp}({\bf p})$ is the distance priors dependent
term which we detail below. 

\subsection{The distance priors}

While SNeIa probe only the background evolution
of the universe over a limited redshift range (up to 
$z \simeq 1.5$), the CMBR anisotropy spectrum and the
matter power spectrum measured by galaxy surveys make it 
possible to both constrain the dynamics of perturbations
and test the model up to the last scattering surface. However,
such a program is both theoretically difficult and computationally
demanding. On the one hand, we must develop the full theory of 
perturbations to compute the ${\cal{C}}_l$ coefficients of the 
multipole expansion of the CMBR anisotropies and the matter power 
spectrum $P(k)$. While this is yet done for the standard dark energy 
models, a full perturbation theory for the vector models we 
are considering is still to be developed. On the other hand, 
even for popular dark energy models, computing both the 
${\cal{C}}_l$ and $P(k)$ represents a bottleneck for the 
algorithms matching data with theories. Fortunately, many features of 
the CMB and matter power spectra may be expressed as a function of a limited subset 
of quantities which are instead easy and straightforward to
compute. Motivated by this consideration, it has become 
popular to summarize the main constraints coming from CMBR and 
matter power spectra in what are defined {\it distance priors}\footnote{
Actually, not all theese quantities are indeed distances. Nevertheless, 
the set of constraints is collectively referred to with this name 
since most of them are easily related to a distance.}. 

In the analysis of the WMAP5 data \cite{WMAP5}, Komatsu et al.
demonstrated that most of the information in the WMAP power spectrum
may be summarized in a set of constraints on the following quantities\,:

\begin{enumerate}

\item[(i.)]{the physical baryon density\,:

\begin{equation}
\omega_b = \Omega_b h^2
\label{eq: defomegab}
\end{equation}
with $\Omega_b$ the density parameter of baryons only;}

\item[(ii.)]{the redshift $z_{LS}$ to the last scattering 
surface that we approximate as \cite{HS96}\,:

\begin{equation}
z_{LS} = 1048 (1 + 0.00124 \omega_b^{-0.738}) [1 + g_1(\omega_b) 
\omega_M^{g_2(\omega_b)}] \ ,
\label{eq: defzls}
\end{equation}

\begin{equation}
g_1(\omega_b) = \frac{0.0738 \omega_b^{-0.238}}{1 + 39.5 \omega_b^{0.763}} \ ,
\label{eq: defg1}
\end{equation}

\begin{equation}
g_2(\omega_b) = \frac{0.560}{1 + 21.1 \omega_b^{1.81}} \ ,
\label{eq: defg2}
\end{equation}
having denoted with $\omega_M = \Omega_M h^2$ the total matter 
physical density;}

\item[(iii.)]{the acoustic scale \cite{larefs}\,:

\begin{equation}
l_A = \frac{\pi (c/H_0) r(z_{LS})}{r_s(z_{LS})} 
\label{eq: defla}
\end{equation}
with $r_s(z_{LS})$ the size of the sound horizon
at the decoupling epoch given by\,:

\begin{equation}
r_s(a_{LS}) = \frac{c/H_0}{\sqrt{3}} 
\int_{0}^{a_{LS}}{\frac{da}{a^2 E(a) \sqrt{1 +  R a}}} \ .
\label{eq: defrs}
\end{equation}
being $a_{LS} = (1 + z_{LS})^{-1}$ and $R = 3\Omega_b/4\Omega_r$;}

\item[(iv.)]{the shift parameter defined as \cite{larefs}\,:

\begin{equation}
{\cal{R}} = \sqrt{\Omega_M} r(z_{LS}) \ .
\label{eq: defshiftpar}
\end{equation}}

\end{enumerate}
While this set of constraints relies on the CMBR data,
none of them has to do with the matter power spectrum. 
To overcome this problem, Eisenstein et al. \cite{Eis05}
introduced the {\it acoustic peak parameter} defined as

\begin{equation}
{\cal{A}} = \frac{\sqrt{\Omega_M H_0^2}}{c z_m} D_V(z_m)
\label{eq: defacpeak}
\end{equation}
where $z_m$ is the median redshift of the galaxy survey 
used to extract the matter power spectrum and the {\it 
volume distance} is given by\,:

\begin{equation}
D_V(z) = \left [ \frac{cz}{H(z)} \ \times
\ \frac{c r^2(z)}{H_0} \right ]^{1/3} \ .
\label{eq: defdv}
\end{equation}
The analysis of the correlation function of a sample 
of more than 46000 LRGs from the SDSS allowed 
Eiseinstein et al. to estimate\,:

\begin{displaymath}
{\cal{A}} = 0.469 \pm 0.017
\end{displaymath}
thus offering another valuable constraint on the model 
parameters\footnote{Note that ${\cal{A}}$ does not depend
on $h$.}. However, a more recent analysis of the seventh 
SDSS data release allowed Sanchez et al. \cite{San09} to 
measure the correlation function along both the radial and 
tangential directions thus allowing a more detailed treatment. 
Adding the CMBR data in a joint analysis, the authors then
provided a new set of distance priors adding to the four quantities
quoted above a new one defined as

\begin{equation}
G(z_m) = r(z_M) \ \times \ [E(z_m)]^{0.8} \ .
\label{eq: defgzm}
\end{equation}
In order to take into account the distance priors, we 
introduce the following likelihood function\,:

\begin{equation}
{\cal{L}}_{dp}({\bf p}) = \frac{1}{(2 \pi)^{5/2} 
|C_{dp}|^{1/2}} \times 
\exp{\left ( -\frac{\Delta_{dp} C_{dp}^{-1} 
\Delta_{dp}^T}{2} \right )} 
\label{eq: deflikedp}
\end{equation}
where $\Delta_{dp}$ is a five dimensional vector whose
i\,-\,th element is given by $\Delta_{dp,i} = x_{i,obs} - 
x_{i,th}$  with the subscript $obs$ and $th$ denoting the 
observed and theoretically predicted values. The label $i$
runs from 1 to 5 referring, respectively, to $\omega_b$, 
$z_{LS}$, $l_A$, ${\cal{R}}$ and $G(z_m)$. Finally, we follow Appendix B of 
\cite{San09} to set the observed values and the covariance 
matrix of the distance priors.

\begin{table}[t]
\begin{center}
\begin{tabular}{ccccc}
\hline
Par $x$ & $\langle x \rangle$ & $x_{med}$ & $68\%$ CL & $95\%$ CL \\
\hline \hline
$\Omega_b$ & 0.0394 & 0.0394 & (0.0385, 0.0404) & (0.0374, 0.0413) \\
$\log{m_{\nu}}$ & -0.036 & -0.037 & (-0.048, -0.025) & (-0.059, -0.012) \\
$h$ & 0.753 & 0.753 & (0.749, 0.757) & (0.745, 0.761) \\
\hline
$\Omega_b$ & 0.0384 & 0.0384 & (0.0372, 0.0394) & (0.0363, 0.0402) \\
$\log{m_{\nu}}$ & 0.005 & 0.006 & (-0.016, 0.025) & (-0.031, 0.041) \\
$h$ & 0.753 & 0.753 & (0.749, 0.758) & (0.745, 0.761) \\
\hline
\end{tabular}
\end{center}
\caption{Summary of the results of the likelihood analysis. The upper 
and lower part of the table refers to the simple and standard MONDian
vector model respectively.}
\end{table}

\subsection{Fitting SNeIa and distance priors}

We now run our MCMC algorithm to explore the models parameters
space now maximimizing the likelihood (\ref{eq: deflikenew}) 
taking care of both the SNeIa and distance priors datasets. 
The median and mean values and 68 and $95\%$ confindence ranges 
are summarized in Table\,III for both the simple and standard MONDian 
vector model. The best fit parameters, i.e. the values of 
$(\Omega_b, \log{m_{\nu}}, h)$ that maximizes the likelihood, 
are not forced to be equal to the  median values because of correlations 
among the parameters. Indeed, we find\,:

\begin{displaymath}
(\Omega_b, \log{m_{\nu}}, h) = (0.0392, -0.038, 0.753)
\end{displaymath}
for the simple MONDian vector model, while it is\,:

\begin{displaymath}
(\Omega_b, \log{m_{\nu}}, h) = (0.0387, -0.001, 0.753)
\end{displaymath}
for the standard case. Both models, however, do not
offer a good performace in fitting the data. Indeed, for 
the best fit parameters, we get\,:

\begin{displaymath}
\chi^2_{SNeIa}/d.o.f = 1.32 \ \ , \ \ 
\omega_b = 0.0222 \ \ , \ \ 
z_{LS} = 1081.7 \ \ ,  
\end{displaymath}

\begin{displaymath}
l_A = 300.0 \ \ , \ \ 
{\cal{R}} = 1.52 \ \ , \ \ 
G(z_m) = 1424 \ \ , 
\end{displaymath}
for the simple case and 

\begin{displaymath}
\chi^2_{SNeIa}/d.o.f = 1.31 \ \ , \ \ 
\omega_b = 0.0219 \ \ , \ \ 
z_{LS} = 1082.3 \ \ ,  
\end{displaymath}

\begin{displaymath}
l_A = 300.4 \ \ , \ \ 
{\cal{R}} = 1.54 \ \ , \ \ 
G(z_m) = 1424 \ \ , 
\end{displaymath}
for the standard model. It is immediately clear from
the unusually high reduced $\chi^2_{SNeIa}$ value signals
that something is going wrong with fitting the SNeIa Hubble 
diagram. Indeed, with $d.o.f. = N_{SNeIa} - 3 = 304$, a 
reduced $\chi^2_{SNeIa}/d.o.f. \simeq 1.3$ has just a tiny 
$\sim 3 \times 10^{-5}$ probability to occur so that we can
safely conclude that both models are not correctly fitting
the SNeIa Hubble diagram data. This conclusion is further enforced comparing the 
above best fit distance priors whose median values and 
standard deviation\footnote{Actually, the standard deviation
is not a good estimator of the uncertainty on the distance 
priors because it does not take into account the 
correlations among them. However, it gives an idea of the
discrepancy between predicted and observed values.} are as follows\,:

\begin{displaymath}
\omega_b = 0.0228 \pm 0.0055 \ , \  
z_{LS} = 1090.1 \pm 0.9 \ , \ 
l_A = 301.6 \pm 0.7 \ ,  
\end{displaymath}

\begin{displaymath}
{\cal{R}} = 1.701 \pm 0.018 \ \ , \ \ 
G(z_m) = 1175 \pm 21 \ \ . 
\end{displaymath}
While the values of $(\omega_b, z_{LS}, l_A)$ are in reasonable 
agreement, there are strong discrepancies for both the shift 
${\cal{R}}$ and the $G(z_m)$ parameters. Considering the $68$ and
$95\%$ confidence ranges in Table IV does not ameliorate the 
comparison so that we must conclude that the model is unable 
to fit both the SNeIa and distance priors dataset. 

\begin{table}[t]
\begin{center}
\begin{tabular}{ccccc}
\hline
Par $x$ & $\langle x \rangle$ & $x_{med}$ & $68\%$ CL & $95\%$ CL \\
\hline \hline
$\omega_b$ & 0.0224 & 0.0223 & (0.0219, 0.0229) & (0.0213, 0.0234) \\
$z_{LS}$ & 1081.6 & 1081.6 & (1081.1, 1082.0) & (1080.7, 1082.6) \\
$l_A$ & 299.9 & 299.9 & (299.2, 300.5) & (298.6, 301.2) \\
${\cal{R}}$ & 1.524 & 1.523 & (1.519, 1.529) & (1.514, 1.533) \\
$G(z_m)$ & 1424 & 1424 & (1416, 1431) & (1407, 1439) \\
\hline
$\omega_b$ & 0.0218 & 0.0218 & (0.0211, 0.0223) & (0.0206, 0.0228) \\
$z_{LS}$ & 1082.6 & 1082.5 & (1081.8, 1083.4) & (1081.2, 1084.0) \\
$l_A$ & 300.6 & 300.6 & (299.9, 301.4) & (299.1, 302.2) \\
${\cal{R}}$ & 1.545 & 1.545 & (1.543, 1.549) & (1.539, 1.551) \\
$G(z_m)$ & 1424 & 1424 & (1416, 1432) & (1409, 1439) \\
\hline
\end{tabular}
\end{center}
\caption{Constraints on the predicted distance priors parameters. The upper
and lower part of the table refers to the simple and standard MONDian
vector model respectively.}
\end{table}

It is worth investigating why this happens. Actually, a hint is
given by noticing that the most discrepant parameters are
those involving $E(z)$ both directly, as for $G(z_m)$, or indirectly 
through an integral, as both ${\cal{R}}$ and $\chi^2_{SNeIa}$. Taking the 
$\Lambda$CDM model as a comparison, we indeed find that both the simple 
and standard MONDian vector models systematically underestimates $E(z)$. 
As a consequence, the dimensionless comoving distance $r(z)$ turns out
to be overestimated thus leading to $\mu(z)$ becoming increasingly higher
than the concordance model prediction as the redshift increases in accordance
with the result that $\mu(z)$ is larger than $\mu_{obs}$ for SNeIa with $z > 1$. 
This also explains why $G(z_m)$ gets larger than observed, while the situation
is different with the shift parameter. Indeed, according to Eq.(\ref{eq: defshiftpar}),
a larger $r(z)$ should lead to a larger ${\cal{R}}$, while we observe the
opposite result. This can, however, be easily explained considering that 
${\cal{R}}$ is also proportional to $\Omega_M$ and we get\,:

\begin{displaymath}
\langle \Omega_M \rangle = 0.091 \ \ , \ \ 
\Omega_{M,med} = 0.091 \ \ ,
\end{displaymath}

\begin{displaymath}
{\rm 68\% \ CL \ :} \ \ (0.089, 0.094) \ \ ,
\end{displaymath}

\begin{displaymath}
{\rm 95\% \ CL \ :} \ \ (0.086, 0.096) \ \ ,
\end{displaymath}
for the simple MONDian vector model and

\begin{displaymath}
\langle \Omega_M \rangle = 0.095 \ \ , \ \ 
\Omega_{M,med} = 0.095 \ \ ,
\end{displaymath}

\begin{displaymath}
{\rm 68\% \ CL \ :} \ \ (0.094, 0.097) \ \ ,
\end{displaymath}

\begin{displaymath}
{\rm 95\% \ CL \ :} \ \ (0.092, 0.098) \ \ ,
\end{displaymath}
for the standard case. These values are much smaller
than the typical $\Omega_M \simeq 0.26$ \cite{D07,K08,WMAP5}
values obtained in literature thus explaining why our
shift parameter turns out to be so small notwithstanding the 
higher $r(z_{LS})$.

\begin{table}[t]
\begin{center}
\begin{tabular}{ccccc}
\hline
Par $x$ & $\langle x \rangle$ & $x_{med}$ & $68\%$ CL & $95\%$ CL \\
\hline \hline
$q_0$ & -0.861 & -0.862 & (-0.865, -0.858) & (-0.869, -0.855) \\
$z_T$  & 1.71 & 1.71 & (1.68, 1.73) & (1.66, 1.76) \\
$t_0$ & 16.93 & 16.93 & (16.84, 17.01) & (16.76, 17.10) \\
\hline
$q_0$ & -0.856 & -0.856 & (-0.859, -0.854) & (-0.861, -0.852) \\
$z_T$  & 1.67 & 1.67 & (1.65, 1.68) & (1.64, 1.70) \\
$t_0$ & 16.77 & 16.77 & (16.66, 16.87) & (16.60, 16.98) \\
\hline
\end{tabular}
\end{center}
\caption{Constraints on derived quantities (with $t_0$ in Gyr). The upper
and lower part of the table refers to the simple and standard MONDian
vector model respectively.}
\end{table}

For completeness, we also summarize in Table V the values 
of the present day deceleration parameter $q_0$, the transition
redshift $z_T$ and the age of the universe $t_0$. These results 
strengthen our conclusion that both models, while performining well 
in the low redshift  regime, are actually quite poor in reproducing 
data probing higher $z$. Indeed, the values of $q_0$ are in reasonable 
agreement with the results quoted in Sect.\,IIIC, even if our $q_0$ 
values are somewhat extreme. On the contrary, there is a clear 
disagreement with the previous estimates of the transition
redshift $z_T$ with our results being very high (also outside
the range probed by SNeIa Hubble diagram). Such a result furtherly 
signals that indeed the expansion rate and hence the transition 
from acceleration to deceleration of our models is too slow. 
Another evidence in favour of this interpretation is provided by 
$t_0$ which turns out to be in strong disagreement with both
the WMAP5 and globular clusters estimate. Indeeed, Eq.(\ref{eq: deftz}) 
shows that, should we underestimate $E(z)$, the age of the universe 
turns out to be higher which is just what happens for both MONDian vector models. 

\subsection{A problem with the models or the data ?}

The large reduced $\chi^2_{SNeIa}$ values, the disagreement between
the best fit predicted and observed distance priors and the 
unacceptably high $z_T$ and $t_0$ results should make us conclude
that both MONDian vector models are unable to fit both the low
and high redshift data. Comparing the results in Table I and III, 
however, makes it evident that the two fitting procedures select
very different regions of the parameter space. It is
therefore worth wondering whether there is a problem with the
distance priors data rather than with the models. 

To this end, it is worth stressing that, although widely used in 
the recent literature (see, e.g., \cite{D07,K08,WMAP5,WM07,dpused}), 
their estimate is  actually model dependent. Indeed, in order to obtain 
their central values and covariance matrix, one first fits
a given model (typically, the concordance $\Lambda$CDM one)
to the full CMBR anisotropy and galaxy power spectra dataset
using a Markov Chain Monte Carlo (MCMC) method to sample the
posterior probability. This same model is then used to compute
the distance priors along the chain and then the sample thus 
obtained is analysed to infer the covariance matrix. As correctly
stressed in \cite{BW09}, this procedure relies on three main 
assumptions\,:

\begin{figure}
\centering \resizebox{8.5cm}{!}{\includegraphics{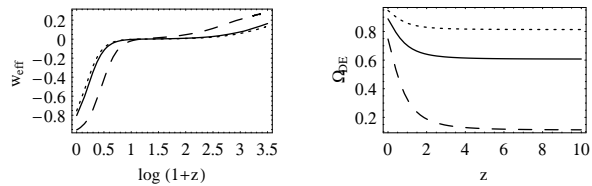}}
\caption{Effective EoS (left) and dark energy density parameter (right) for the 
standard MONDian vector model. We set $(\Omega_b, h) = (0.04, 0.70)$ 
and consider three values for $\log{m_{\nu}}$, namely -1.0 (short dashed),
$0.0$ (solid) and 0.5 (long dashed).}
\label{fig: standde}
\end{figure}

\begin{enumerate}

\item{the posterior probability from the CMBR anisotropies and galaxy 
power spectra is correctly described by the distance priors, i.e. no 
information is lost when giving away the full dataset in favour of the 
simplified one;}

\item{the mean and covariance matrix of the distance priors parameters
do not change when the model space is enlarged, i.e. the choice of the 
fiducial model does not impact the estimate of the priors;}

\item{these summary parameters are weakly correlated with the input 
fiducial model parameters so that they can be used as independent constraints.}

\end{enumerate}
While the first and third assumptions have indeed been 
verified \cite{WM07,BW09}, the second hypothesis can not 
be fully addressed. On the one hand, one could argue that, unless
a significative amount of dark energy in the early  universe is present, 
all reliable models should match at high redshift, when 
the effect of dark energy fades away. As a consequence, using the 
$\Lambda$CDM as a fiducial model to compute quantities that mainly 
depend on the high redshift behaviour should not affect the 
final estimate. However, even if the background evolution could be 
the same, it is still possible that the dynamics of perturbation
is radically different as is the case with modified gravity theories.
This is indeed what also happens for the MONDian vector models we are
considering. Fixing the model parameters to the best fit values from
the fit to the low redshift data only, we can easily compare the 
luminosity distance with that for a fiducial $\Lambda$CDM scenario. 
As expected (since they both reproduce the same data), the two can hardly
be discriminated over the range probed, while the Hubble parameter
only differ by a few percent. While the background evolution is therefore
comparable, perturbations evolve in a completely different way because
of the presence of the vector field. As such, one can not be sure 
that the information contained in the CMBR spectrum may still be 
summarized in the distance priors quantities as for the $\Lambda$CDM 
model. A similar discussion also applies for the the prior on $G(z_m)$. Moreover, 
it is worth stressing that this latter quantity actually depends on data 
probing an intermediate redshift range so that one can not rely anymore on
the fading of dark energy at high $z$ as is typically done for quantities 
as, e.g., the acoustic scale $l_A$ and the shift parameter ${\cal{R}}$. 

Investigating the impact of these problems on the estimate of the 
distance priors is outside our scope here. We nevertheless stress
that, because of the above considerations, one can not safely reject
the MONDian vector models because of the disagreement with  the
distance priors values. As a conservative conclusion, we are 
therefore forced to only report the results being unable to decide 
whether the problem is with the data or with the fiducial model used
to retrieve them.

\subsection{The effective EoS and the high $z$ limit}

An alternative way to compare the proposed MONDian vector
models with standard dark energy models in both the low and 
high redshift regime may be obtained by considering the effective
EoS. Indeed, from the point of view of the background evolution, our
models are equivalent to a cosmological scenario made out of 
dust matter and a dark energy with an effective EoS given by\,:

\begin{eqnarray}
1 + w_{eff}(z) & = & \left [ \frac{2}{3} \frac{d\ln{E(z)}}{d\ln{(1 + z)}} - 
\frac{\Omega_M (1 + z)^3}{E^{2}(z)} \right ] \nonumber \\
~ & \times & \left [ 1 - \frac{\Omega_M (1 + z)^3}{E^{2}(z)} \right ]^{-1} \ ,
\label{eq: eoseff}
\end{eqnarray}
so that the dark energy density parameter reads\,:

\begin{figure}
\centering \resizebox{8.5cm}{!}{\includegraphics{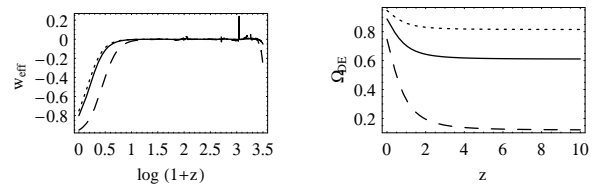}}
\caption{Same as Fig.\,\ref{fig: standde}, but for the simple MONDian vector model. 
The small ripples are due only to numerical errors.}
\label{fig: simplede}
\end{figure}

\begin{equation}
\Omega_{DE}(z) = \frac{1 - \Omega_M}{E^2(z)} 
\exp{\left [ 3 \int_{0}^{z}{\frac{1 + w_{eff}(z')}{1 + z'} dz'} \right ]} \ .
\label{eq: omdeeff}
\end{equation}
Figs.\,\ref{fig: standde} and \ref{fig: simplede} show $w_{eff}(z)$ and 
$\Omega_{DE}(z)$ for the standard and simple MONDian vector models setting
the baryon density parameter to the fiducial value $\Omega_b = 0.04$ and
varying the neutrino mass $m_{\nu}$. Note that the value of $\Omega_M$ in
Eqs.(\ref{eq: eoseff}) and (\ref{eq: omdeeff}) is related to $\Omega_b$ and 
$m_{\nu}$ through Eq.(\ref{eq: omvsobmnu}) so that changing $m_{\nu}$ is the 
same as varying $\Omega_M$. 

Both figures shows that the value of $m_{\nu}$ set the present day EoS with 
$w_{eff}(z = 0)$ increasing as a function of $m_{\nu}$, even if there is a 
saturation at small $m_{\nu}$ as can be easily understood considering that the
smaller is $m_{\nu}$, the less is the impact of massive neutrinos in the energy
budget. It is worth noting that for $\log{m_{\nu}} \simeq 0.5$, it is $w_{eff}(z = 0)
\simeq -1.0$ and $dw_{eff}/dz$ takes very small values over the redshift range $(0, 1)$, 
i.e. we recover (for both cases) a present day cosmological constant. It is therefore 
not surprising that the fit to the SNeIa dataset points towards such values of 
$\log{m_{\nu}}$ since these make both MONDian vector models as similar as possible 
to the $\Lambda$CDM one over the range probed by SNeIa themselves. 

On the contrary, in the high redshift regime, the neutrino mass only plays a marginal
role in determining the value of $w_{eff}(z)$ which stays almost constant at value 
close to the dust one $w_{eff} = 0$, i.e. the effective EoS approximately tracks the 
matter term. However, the amount of dark energy is different depending on $m_{\nu}$ with
lower values of the neutrino density giving rise to a larger $\Omega_{DE}(z)$ in the 
high $z$ regime. Such a behaviour is somewhat counterintuitive since one expects that
the effective dark energy fades away in the early universe in order to recover the matter
dominated epoch. Actually, one must take into account that our matter term is made out of 
baryons and massive neutrinos only. The lower is the neutrino mass, the higher must be 
the contribute of the effective dark energy (that in this regime behaves as matter being
$w_{eff} \simeq 0$) to compensate for the missing cold dark matter. Such a result can also
be forecasted going back to Eq.(\ref{eq: f1}) and noting that, for our best fit models,
we find $\eta >> 1$ so that, for $z >> 1$, it is ${\cal{K}} >> 1$ too. Let us then 
consider the simple MONDian vector model. Inserting Eq.(\ref{eq: fksimple}) and taking
the limit for $\eta E(z) >> 1$, we approximately get\,:

\begin{displaymath}
\Omega_M (1 + z)^3 \simeq E^2 + \ln{\eta E} \simeq E^2
\end{displaymath}
so that we indeed recover the usual expression for the Hubble parameter for a 
matter dominated universe with $\Omega_M$ the total (baryons and massive neutrinos) 
matter term (having neglected radiation). From the point of view of the effective
dark energy formalism, if we reduce the neutrinos contribution, we must add a further 
term acting as matter and this is indeed provide by the effective dark energy. This 
also explains why $\Omega_{DE}(z >> 1)$ increases decreasing $\log{m_{\nu}}$. A similar
discussion also applies to the standard MONDian vector models so that in both cases    
the final scenario is the one of a universe where the matter term 
disappears to be replaced by a dark energy fluid acting approximately as matter. This 
explains why the inclusion of the distance priors (probing the high $z$ regime) pushes 
the best fit towards smaller $\log{m_{\nu}}$ since, in this case, both MONDian models 
recovers the usual Friedmann models which are known to successfully fit these high 
redshift probes. As yet said, however, decreasing $\log{m_{\nu}}$ makes $w_{eff}(z = 0) 
\ne -1$ thus worsening the fit to the SNeIa Hubble diagram.

\section{Conclusions}

The astonishing successes of MOND on 
the galactic scales and the emergence of a 
relativistic theory playing the role of its
counterpart on cosmological scales have renewed
the interest in the search for a possible common
explanation of both dark matter and dark energy 
phenomenology. Vector theories are another way 
of recovering MOND in the low energy limit so that
it is worth wondering whether they can also offer 
an elegant way of speeding up the cosmic evolution
without the need of any dark energy source. The ignorance 
about the form of the vector field Lagrangian may be 
bypassed relying on the link between the function 
${\cal{F}}({\cal{K}})$ and the MOND interpolating 
function $\mu(a/a_0)$. Since dark matter is no more 
present, one should postulate the presence of massive
neutrinos in order to fill the gap between the total matter
density parameter and the baryons density alone. Moreover, 
such massive neutrinos are also advocated in order to
reconcile the results of solar and atmospheric neutrino
experiments on the flavour mixing with the predictions of 
the standard model of particle physics. 

Motivated by these considerations, we have therefore
investigated the viability of two different MONDian vector
models characterized by ${\cal{F}}({\cal{K}})$ expressions
corresponding to the simple and standard MOND interpolating 
function. To this aim, we have first fitted them against the
Union SNeIa Hubble diagram using Gaussian priors on the physical
baryon density $\omega_b$ and the present day Hubble constant $h$
in order to break the $(\Omega_b, \log{m_{\nu}})$ degeneracy. 
Both models performs quite well giving a perfect agreement with 
the SNeIa data and previous  estimates of the total matter density 
parameter $\Omega_M$, the deceleration parameter $q_0$ and the age 
of the universe $t_0$. Moreover, the (weak) constraints on the neutrino 
mass are consistent with the upper limits set by the Mainz\,-\,Troitz 
experiment assuming three families of degenerate neutrinos. Should, instead,
a single sterile neutrino be the mass dominant component, its estimated mass is 
in agreement with the constraints from the Miniboone experiment
also falling in the right range advocated to solve MOND problems on 
cluster scales. 

In order to investigate the high redshift behaviour of the models,
we have repeated the likelihood analysis adding the extended set of distance
priors. It turns out that both models should be rejected since 
they provide now a poor fit to the SNeIa data and strongly 
disagree with the observed shift ${\cal{R}}$ and $G(z_m)$ 
parameters. Distance priors are, however, estimated through 
a model dependent procedure so that one can not safely rely
on them to exclude models that are radically different from
the fiducial one used to extract the constraints on ${\cal{R}}$
and $G(z_m)$. As a consequence, we are unable to conclude whether
the disagreement is a failure of the MONDian vector models or
an expected outcome of the different evolution of perturbations
with respect to usual dark energy models because of the action
of the vector field.

It is worth remembering that the search for a cosmological
counterpart to MOND has a long history with vector models 
being only the most recent proposal. Applying a procedure which
is simply the generalization of the classical derivation of the 
Friedmann equation from the Newtonian force law, but now starting
from the MOND force law, Lue \& Starkman \cite{LS04} derived an
expression for the Hubble parameter without referring to any
underlying modified gravity theory. Defining $g(x) = H^2/H_0^2$ and 
$x = \Omega_M (1 + z)^3$, they indeed find\,:

\begin{equation}
g(x) = \left \{
\begin{array}{ll}
(\beta \ln{x} + c_2) x^{2/3} & x < x_c \\
~ & ~ \\
x + c_1 x^{2/3} & x > x_c \\
\end{array}
\right . 
\label{eq: lsgx}
\end{equation} 
with $c_1 = c_2 + 3\beta [\ln{(3\beta)} - 1]$. Moroever, 
they also proposed a modified (by hand) version of this expression 
to better account for the low redshift behaviour\,:

\begin{figure}
\centering \resizebox{8.5cm}{!}{\includegraphics{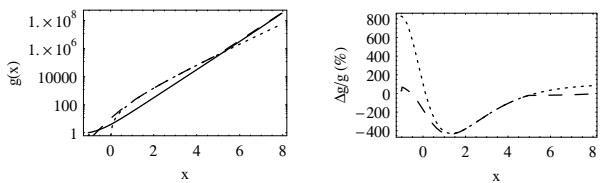}}
\caption{{\it Left\,:} the $g(x)$ function for the best fit standard 
MONDian vector model (solid line) compared to the Lue \& Starkman proposals.  
{\it Right\,:} relative deviation of the proposal $g(x)$ from that of our model. 
Short and long dashed lines refer to $g(x)$ given by Eqs.(\ref{eq: lsgx}) and 
(\ref{eq: lsgxbis}) respectively.}
\label{fig: lsfit}
\end{figure}

\begin{equation}
g(x) = \left \{
\begin{array}{ll}
\Omega_{\Lambda} & x \le 0.1 \\
~ & ~ \\
\beta x^{2/3} \ln{(1 + z)} & 0.1 \le x \le (3 \beta)^3 \\
~ & ~ \\
x + 3 \beta [\ln{(3 \beta)} - 1] x^{2/3} & x > (3 \beta)^3 \\
\end{array}
\right . \ .
\label{eq: lsgxbis}
\end{equation} 
Following \cite{LS04}, we set $(\beta, x_c, \Omega_{\Lambda}) = 
(15, 7 \times 10^{4}, 0.7)$ with $c_2 = 0$ and show in Figs.\,\ref{fig: lsfit}
and \ref{fig: lsfitbis} how Eqs.(\ref{eq: lsgx}) and (\ref{eq: lsgxbis}) compares
to the standard and simple MONDian vector models. As it is better appreciated
from the right panels, both Eqs.(\ref{eq: lsgx}) and (\ref{eq: lsgxbis}) 
agree reasonably well with the $g(x)$ expression of our MONDian models for 
very high $x$, that is in the early universe. This is an expected result since,
in this redshift range, both our models and the phenomenological Lue \& Starkman 
proposals recover the typical General Relativity matter dominated scenario. 
On the contrary, for small $x$, i.e. in the very low $z$ regime, only the
modified expression (\ref{eq: lsgxbis}) matches reasonably well those
for the standard and simple MONDian vector models as a consequence of 
both mimicking an effective cosmological constant. In the intermediate
region, however, matching the Lue \& Starkman models to our own is impossible 
signalling that the strategy adopted by these authors is unable to trace
the transition region to General Relativity. However, it is also worth 
stressing that such a disagreement could be expected since the Lue \& Starkman
procedure relies on the assumption that, whatever the underlying modified
gravity theory leading to (\ref{eq: lsgx}) or (\ref{eq: lsgxbis}) is, the Birkhoff
theorem still holds. This is not the case for vector theories \cite{noBirk} 
so that the two approaches differ from the very beginning.

\begin{figure}[t]
\centering \resizebox{8.5cm}{!}{\includegraphics{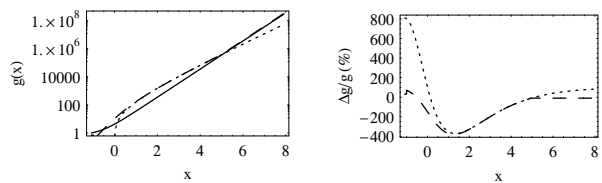}}
\caption{Same as Fig.\,\ref{fig: lsfit} but for the simple model.}
\label{fig: lsfitbis}
\end{figure}

As yet quoted in the introduction, much interest has been devoted 
to the TeVeS theory as a relativistic MOND counterpart. In particular, 
Skordis et al. \cite{Skordis} has also investigated the 
growth of structure in TeVeS also computing the CMBR anisotropy
spectrum. It turned out that the inclusion of a cosmological constant term 
and of massive neutrinos with $m_{\nu} \simeq 2 \ {\rm eV}$ may lead to a 
reasonably good agreement with the data, although a detailed fitting has 
not been performed. Since both the vector and scalar fields present in TeVeS 
do not contribute significantly to the dynamics, it is likely that this 
TeVeS\,+\,$\Lambda$ model matches well both the SNeIa Hubble diagram and the 
distance priors. However, comparing the Skordis et al. model to our MONDian
vector theories is not possible given the radical differences betwee the 
two approaches. Indeed, TeVeS needs the scalar field to act as a dark matter like
term on galactic scales, while the vector field boosts the growth of 
perturbations during the $\Lambda$ driven background expansion. On the contrary,
in our approach, the vector models modify the low energy limit Poisson equation
thus recovering the MOND\,-\,like behaviour, but also originates the cosmic speed up. 
In a sense, our approach is more {\it economical} claiming for the lowest 
possible number of ingredients. 

The positive results obtained in the low redshift regime 
may be considered only as a ground zero level analysis. 
Indeed, fitting the SNeIa Hubble diagram only tells us that 
the two considered MONDian vector models predict
the correct background evolution over the redshift range probed
by the data, i.e. up to $z \sim 1.5$. Needless to say, more tests
are needed in order to assess the viability of these models. On one 
hand, we can still investigate the background evolution by extending
the redshift range through the use of the GRBs Hubble diagram \cite{GRBhd}.
On the other hand, a more significative and demanding test should be 
fitting both the galaxy power spectrum and the CMBR dataset. However,
both these tasks are quite daunting from the theoretical point of view. 
Indeed, as far as we know, the CMBR spectrum has never been computed for
the class of vector theories we are considering so that one should first
write down the full set of perturbation equations to then modify a 
numerical code like CAMB \cite{CAMB}. Somewhat easier (even if still
difficult) is to deal with the problem of growth of structures in 
vector models. Indeed, such a study has although been performed in \cite{ZFSbis}
where the authors considered a simple power\,-\,law choice for 
${\cal{F}}({\cal{K}})$. Although a detailed fitting to the data was not
performed, these authors have convicingly shown that the new degrees of 
freedom sourced by the vector field may indeed boost the growth of structure 
even in absence of any dark matter. This result is a good starting point 
for testing our proposed MONDian vector models provided one accordingly 
changes the ${\cal{F}}({\cal{K}})$ function and introduces massive neutrinos
into the game. Note that, since we want to recover MOND on galactic scales, 
we are postulating no cold dark matter so that the observed galaxy power 
spectrum should be matched to the predicted one with a bias parameter 
determined by the clustering properties of the massive (sterile or not) neutrino. 
Therefore, such a test is particularly powerful and worth 
addressing as the next step of our analysis.

It is worth noting that help investigating the viability of our 
MONDian vector models may come from a non astrophysical experiment. 
Indeed, one of the key ingredients in both MONDian vector models is the 
presence of massive neutrinos with $m_{\nu} \sim 2 \ {\rm eV}$. The KATRIN
experiment \cite{KATRIN} on the tritium $\beta$ decay should be able 
to constrain the electron neutrino mass with a sensitity of $\sim 0.2 \ {\rm eV}$.
Should this experiment indeed find that neutrinos are less massive than, e.g., 
$\sim 1 \ {\rm eV}$, our model could be in serious trouble unless 
one assumes that a single massive sterile neutrino does indeed exist.

As a final remark, one could note that we have titled our paper
with a question so that the reader could now claim for an 
answer. Unfortunately, because of the uncertainties on the use
of the distance priors and the small redshift range probed, the 
unique answer we can give after our analysis is only a (somewhat frustrating) {\it maybe}.

\acknowledgments 

It is a pleaure to thank A. S\'anchez for making
available the covariance matrix of the distance priors 
in electronic form and for the illuminating comments on their use. 
We also warmly thank G. Angus for the instructive discussion
on sterile neutrinos in MOND, and A. Diaferio and A. Tartaglia for
a careful reading of the manuscript. VFC is supported by University of 
Torino and Regione Piemonte. Partial support from INFN project PD51 
is acknowledged too.

\end{document}